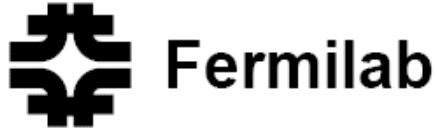

# ADVANCED BENT CRYSTAL COLLIMATION STUDIES AT THE TEVATRON (T-980)[*][†]

V. Zvoda[#], G. Annala, R. Carrigan, A. Drozhdin, T. Johnson, S. Kwan,
N. Mokhov, A. Prosser, R. Reilly, R. Rivera, L. Uplegger, V. Shiltsev, D. Still, J. Zagel,
Fermilab, Batavia, IL 60510, U.S.A.
V. Guidi, E. Bagli, A. Mazzolari, INFN-Ferrara, Italy
Yu. Ivanov, PNPI, Gatchina, Leningrad Region, RU-188300, Russia
Yu. Chesnokov, I. Yazynin, IHEP, Protvino, Moscow Region, RU-142284, Russia

## Abstract

The T-980 bent crystal collimation experiment at the Tevatron [1, 2] has recently acquired substantial enhancements. First, two new crystals - a 16-strip one manufactured and characterized by the INFN Ferrara group and a quasi-mosaic crystal manufactured and characterized by the PNPI group. Second, a two plane telescope with 3 high-resolution pixel detectors per plane along with corresponding mechanics, electronics, control and software has been manufactured, tested and installed in the E0 crystal region. The purpose of the pixel telescope is to measure and image channeled (CH), volume-reflected (VR) and multiple volume-reflected (MVR) beam profiles produced by bent crystals. Third, an ORIGIN-based system has been developed for thorough analysis of experimental and simulation data. Results of analysis are presented for different types of crystals used from 2005 to present for channeling and volume reflection including pioneering tests of two-plane crystal collimation at the collider, all in comparison with detailed simulations.

[*]Work supported by Fermi Research Alliance, LLC under contract No. DE-AC02-07CH11359 with the U.S. Department of Energy through the US LARP Program.
[†]Presented paper at Particle Accelerator Conference'11, March 28 – April 1, 2011, New York, U.S.A.
#zvoda@fnal.gov

# Advanced Bent Crystal Collimation Studies at the Tevatron (T-980)*

V. Zvoda#, G. Annala, R. Carrigan, A. Drozhdin, T. Johnson, S. Kwan,
N. Mokhov, A. Prosser, R. Reilly, R. Rivera, L.Uplegger, V. Shiltsev, D. Still, J. Zagel,
FNAL, Batavia, IL 60510, U.S.A.
V. Guidi, E. Bagli, A. Mazzolari, INFN-Ferrara, Italy
Yu. Ivanov, PNPI, Gatchina, Leningrad Region, RU-188300, Russia
Yu. Chesnokov, I. Yazynin, IHEP, Protvino, Moscow Region, RU-142284, Russia

*Abstract*

The T-980 bent crystal collimation experiment at the Tevatron [1, 2] has recently acquired substantial enhancements. First, two new crystals - a 16-strip one manufactured and characterized by the INFN Ferrara group and a quasi-mosaic crystal manufactured and characterized by the PNPI group. Second, a two plane telescope with 3 high-resolution pixel detectors per plane along with corresponding mechanics, electronics, control and software has been manufactured, tested and installed in the E0 crystal region. The purpose of the pixel telescope is to measure and image channeled (CH), volume-reflected (VR) and multiple volume-reflected (MVR) beam profiles produced by bent crystals. Third, an ORIGIN-based system has been developed for thorough analysis of experimental and simulation data. Results of analysis are presented for different types of crystals used from 2005 to present for channeling and volume reflection including pioneering tests of two-plane crystal collimation at the collider, all in comparison with detailed simulations.

## INTRODUCTION

A bent crystal, used in an accelerator collimation system, can coherently direct channeled halo particles deeper into a nearby secondary absorber. This results in a reduction of outscattering from the system, thereby reducing beam losses in critical locations and radiation loads to the downstream superconducting magnets. Besides channelling process in the bent crystal, the reflection of the particles from bent crystalline planes takes place when particles pass through the crystal or through the array of the aligned crystals (case of multiple volume reflection). That is why beginning from 1980s collimation systems based on bent crystals are considered as an interesting promising option for high-energy accelerators and colliders.

## THE TEVATRON EXPERIMENT

The Tevatron T-980 experiment layout [1, 2] is illustrated in Fig. 1. Current setup incorporates two goniometers, vertical and horizontal, supporting the use of 3 crystals (2 in the vertical plane and 1 in the horizontal). The horizontal goniometer is mounted with an O-shaped crystal O-05-09. The vertical goniometer is mounted with a multistrip MS-08-09 crystal in the upstream position and a multistrip crystal MS-16-11 in the downstream position. Both goniometers are located ~ 27m (vertical) and 24m (horizontal) upstream of a secondary collimator E03. Each crystal is inserted into the beam halo at $5\sigma$ from the core and the angle of the crystal is changed to locate angular position for CH, VR and MVR. The E03 collimator intercepts channelled beam for O-shaped and quasi-mosaic crystals and MVR beam for multistrip crystals.

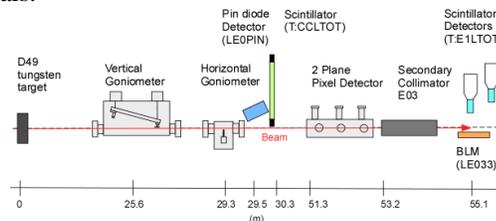

Figure 1: E0 layout.

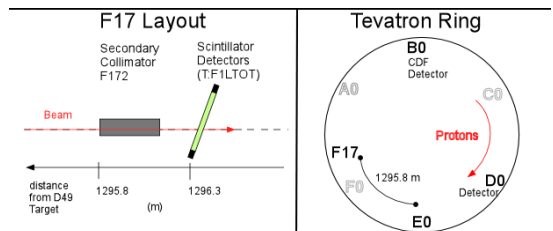

Figure 2: F17 layout and its position at the ring.

The F17 collimator intercepts VR beam for O-shaped and quasi-mosaic crystals and channelled beam for multistrip crystals. The F17 collimator is located about 1.3 km downstream of E0 (see Fig. 2).

*T-980 Crystals*

Three types of bent crystals were used during the experiments existence 2005 to present: O-shaped, multistrip and quasi-mosaic crystal. Below is a short description, the main parameters and the schematic pictures of each type of crystal. Table 1 gives an overview of all crystals used for the experiment T-980.

**O-shaped**. There are two silicon O-shaped crystals: O-BNL-02 and O-05-09, which utilize the (110) channelling plane. The bend is achieved with the stainless steel holder by applying inward pressure on the long sides of the crystal, forcing the short side to bulge outward (see Fig. 3). O-BNL-02 was used as a horizontal deflector for about 4 years (2005-2009) and it has 410 $\mu rad$ bend angle, 5 mm size along the beam direction and 1.6 $mrad$ miscut angle. O-05-09 has 360 $\mu rad$ bend angle, 5 mm length along the beam and the miscut angle of

___________________
*Work supported by Fermi Research Alliance, LLC under Contract DE-AC02-07CH11359 with the U.S. DOE through the US LARP.
#zvoda@fnal.gov

Table 1: Crystals used for studies

| Name of the crystal | CH plane, deflection type | Origin | Bend angle/VR angle, urad | Size along the beam, mm | Critical angle (CH), urad | Miscut angle, urad | Displacement of particles at E03 (CH/VR), mm, simulated | Displacement of particles at F172 (CH/VR), mm, simulated | Maximum Efficiency, % |
|---|---|---|---|---|---|---|---|---|---|
| O-BNL-02 | (110), hor | PNPI | 410/9 | 5 | 5.34 | 1600 | -10/0.43 | 25.55/-1.13 | CH 75 |
| O-05-09 | (110), hor | PNPI | 360/16 | 5 | 5.45 | 120 | -8.13/0.39 | 22.44/-1.06 | CH 60 |
| MS-08-09 | (111), vert | IHEP | 200 and 8/strip | 0.9 | 3.81 | small | 5.04/-1.56 | -6.58/2.07 | VR 70 |
| MS-16-11 | (110), vert | INFN | 250 and 13.5/strip | 1/strip | 3.5 | 600 | 6.3/-5.36 | -8.23/7.11 | |
| QM-01-10 | (111), vert | PNPI | 120/15 | 2 | 5.58 | 50 | 2.9/-0.37 | -3.95/0.49 | |

$120\ \mu rad$. It is currently installed at the Tevatron in the horizontal plane.

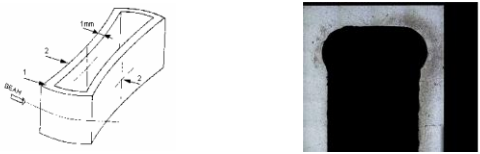

Figure 3: Schematic illustration of O-shaped crystal (left) and the microscope image of O-BNL-02 (right).

The microscope plot (see Fig. 3) shows the average roughness of the surface to be about 10 microns.

**Multistrip**. Fig. 4 shows a multi-strip crystal MS-08-09 containing 8 strips and the curvature scheme for every strip: a mechanical holder bends the strips along the major direction producing a primary curvature (indicated with $P_c$); the anticlastic forces generate a secondary curvature (indicated with $A_c$) which is used to deflect the proton beam. The size of every strip along the beam is 0.9 mm. The bending angle is $200\ \mu rad$ and the angle for multiple volume reflection is $\vartheta_{MVR} \approx 8 \times 8 = 64\ \mu rad$.

A new multistrip crystal, MS-16-11, with 1mm per strip size along the beam was installed in January 2011. Bending angle is $250\ \mu rad$ and $\vartheta_{MVR} \approx 13.5 \times 16 = 216\ \mu rad$. This crystal is still waiting for beam tests.

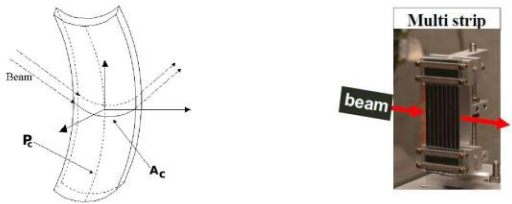

Figure 4: Curvature scheme of strip crystal (left) and photo of MS-08-09 (right).

**Quasi-Mosaic.** This crystal exploits the elastic quasi-mosaicity effect. The bending device again uses anticlastic effects which are designed to bend the crystal in the one plane conforming it to the principal curvature. The anticlastic forces produce a secondary curvature in the perpendicular plane which causes the quasi-mosaic curvature of the atomic plane (in our case (111)).

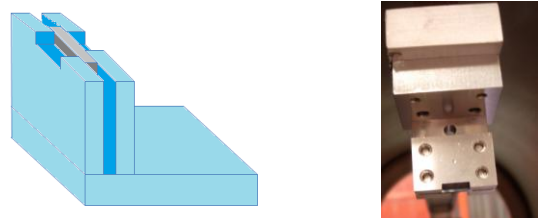

Figure 5: Schematic plot of Quasi-mosaic crystal (left) and photo of QM-01-10 (right).

The size along the beam of QM-01-10 is 2 mm (see Fig. 5) and the window for the incident beam is $2 \times 10$ mm. The bend angle is $120\ \mu rad$ and the miscut angle is $50\ \mu rad$.

*Angular and collimator scans*

There are two types of scans produced during End-of-Store studies for each crystal. The results allow the channelling efficiency and the collimation efficiency to be estimated for the crystal and the system as a whole. Angular scans are performed by measuring losses at E03 while the orientation of the crystal angle is changed (see Fig. 6 for O-BNL-02). The E03 collimator distance from the beam is set to roughly 6-7σ, slightly further than that of the crystal. In principle, a loss peak is expected on the E03 collimator when the halo particles are aligned with the crystalline planes at the entrance face of the crystal. The angular acceptance for channelling is predicted to be $1.5\vartheta_c \approx 10\ \mu rad$, where $\vartheta_c \approx 6\ \mu rad$.

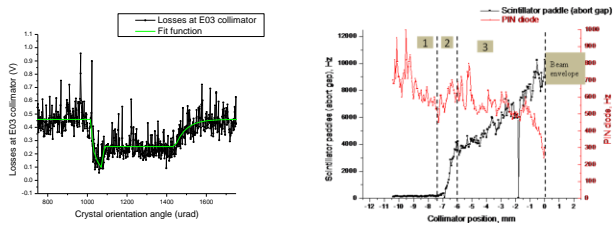

Figure 6: Typical angular and collimator scans.

Collimator scans are used to measure the deflection of the channeled beam once the crystal angle is set to the channeling peak. The E03 collimator is slowly being moved into the beam starting from a completely retracted position.

There are three distinct regions (see Fig. 6):

1. A region of negligible losses, where the collimator does not intercept any beam.
2. A steep increase in the losses, where the collimator intercepts the channeled beam.
3. A region where the losses increase slowly as the collimator begins intercepting dechannelled and amorphous scattered particles.

When the collimator finally touches the beam envelope, the losses rise abruptly and the PIN diode signal decreases to zero, as the collimator becomes the primary scatterer masking the crystal. For O-BNL-02, the expected displacement of the channeled beam (middle of Region 2) with respect to the beam envelope is 10 mm, while the measured distance is only 7 mm. For O-05-09 and for MS-08-09 the expected displacements of CH beam are about 8 mm and 7 mm, respectively.

*Analysis system (Origin based)*

The data analysis based on an Origin Software package has been developed to analyze all T-980 data. For angular scan data, a piece-wised fitting function consists of 2 exponential parts, a Gaussian function corresponding to the channelling peak, and a plateau region corresponding to volume reflection or multi volume reflection. For collimator scans, the fitting function consists of the error function for the channelling region, a quadratic function for the dechannelled particles and an exponential tale for the amorphous scattering region. Also, the baseline (constant value) corresponding to the background noise is added to this function. The result allows an estimate of the efficiency of the channelling process in the crystal, as well as determining important parameters such as the channelling peak angle, CH and VR angle acceptance, etc., as shown in Table 1.

*PIXEL telescope*

A PIXEL telescope has been designed and built for observation and tracking of deflected particles in front of the E03 collimator. The design of the E0 pixel is based on the original CMS forward pixel detector (see Fig. 7). It allows tracking of particles through the 3 pixels providing a profile or image of the channelled or volume reflected particles.

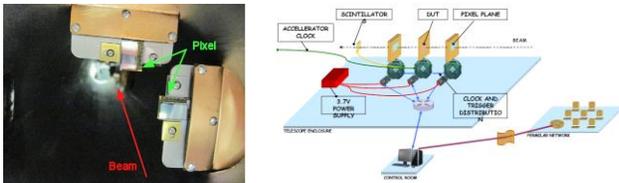

Figure 7: One set of horizontal and vertical pixel detectors (left) and one plane setup for the PIXEL (right).

Each pixel chip of size $1 \times 2 cm^2$ (see Fig. 7) has a sensitive area $0.8 \times 1.6 cm^2$. The resolution of the pixel is 7-8 $\mu m$. Each pixel has size of $100 \times 150 \mu m^2$.

The PIXEL telescope is currently installed at E0. The CAPTAN data acquisition system has been developed by the Detector Instrumentation Group of Computer Division, Electronic Systems Engineering, FNAL [3] and it is supported by Ethernet networking.

## CONCLUSIONS AND PLANS

The T-980 group has achieved a few significant results over last 6 years. Both O-shaped crystals are easy to operate; they produce clear reproducible signals for CH and VR regions. Also, they have a significant bend angle, so the deflected channeled and volume reflected particles could be intercepted by the E03 and F17 collimators. The estimated channeling efficiency is about 75 % for O-BNL-02 and 70 % for O-05-09.

Multistrip crystals have a broad angular acceptance signifying their biggest advantage for operation and performance in the Tevatron. MS-08-09 demonstrated both CH and MVR effects with estimated volume reflection efficiency of about 70%.

The quasi-mosaic crystal has been studied for the past few months. However, it did not produce any relevant channeling effects. After a detailed analysis, it has been concluded that the QM-PIN diode configuration needed improvements and, most important, the beam was hitting the boundary of the crystal and holder. This mainly stems from the style of holder the crystal is mounted in and beam conditions at the end of the colliding beam stores. Constraining the crystal in the holder is too difficult to provide reproducible results in an operational collider [4].

Results and experience of the T-980 experiment are important to draw conclusions and make recommendations on the use of bent crystal systems in the energy frontier colliders, including LHC. One of them is that T980 certainly favors the use of the multistrip crystals. The experiment will continue until the end of the Tevatron Collider Run II (which is currently scheduled for Oct 1, 2011), with focus on imaging channeled, volume reflected and multiple-VR beams with the E0 pixel telescope. We also plan to demonstrate an operational effectiveness of a 2-plane crystal collimation system.

## ACKNOWLEDGMENTS

We gratefully acknowledge help from A. Dzyuba, L. Cooley, V. Previtali and A. Vlasov for their contributions over the course of the T-980 experiment, and the H8RD22/UA2 project for support of the new crystal designs.